\documentstyle[aps,epsf,preprint]{revtex}

\begin{document}
\draft
\preprint{HEP/123-qed}
\title{Effect of time-correlation of input patterns on the convergence
of on-line learning}
\author{Tsuyoshi Hondou and Mitsuaki Yamamoto}
\address{
Graduate School of Information Sciences\\
Tohoku University, Sendai 980-77, Japan
\\
}
\author{
 Yasuji Sawada and  Yoshihiro Hayakawa\\
}
\address{
Research Institute of Electrical Communication \\
Tohoku University, Sendai 980-77, Japan
\\
(Received 17 May 1995)
}
\maketitle
\begin{abstract}
We studied the effects of time-correlation of subsequent patterns 
on the convergence of on-line learning by a feedforward neural network with 
backpropagation algorithm.
By using chaotic time series as sequences of 
correlated patterns,
we found that the unexpected scaling of converging time with learning 
parameter emerges 
when time-correlated patterns accelerate learning process.
\end{abstract}

 It has been reported\cite{Mpitosos,Prog,Nakaj} that time-correlation of
 input patterns often largely
influences the convergence of on-line learning.
As a concrete example, 
 learning of chaotic map was shown to converge  
 faster when patterns
appeared  in deterministic order of chaos than 
 when patterns appeared
randomly with 
 the same 'probability 
density' with the chaotic time series
\cite{Mpitosos,Prog}.
 This showed that on-line learning is sensitive to the 
order of subsequent patterns. 
But the influence of the time-correlation on the convergence of on-line
learning has not been
analyzed yet.

 If we express the input and output as vectors, supervised learning is 
a task to acquire the
 mapping relation: $\vec{X}_{p} ( \in  {\bf R^{n}}) \mapsto \vec{Y}_{p} ( \in
  {\bf R^{n}}) $ 
($p \in {\bf N}$ or ${\bf R}$)
 where the set $\{ \vec{X}_{p}, \vec{Y}_{p} \}$ is called 'pattern', 
and $p$ is a pattern index.
When the pattern index, $p$, is continuous, the number of patterns, $L$, is
 infinite. 
In gradient descent learning algorithms, 
the neural network system is updated as follows:
\begin{equation}
\left\{
\begin{array}{rcl}
\vec{\omega}_{n+1} & = & \vec{\omega}_{n}+\delta\vec{\omega}_{n} \\
\delta\vec{\omega}_{n} & = & -\epsilon \nabla_{\vec{\omega}}
E_{n}|_{\vec{\omega}=\vec{\omega}_{n}} ,
\end{array}
\right.
\label{BP}
\end{equation}
where $\vec{\omega}_{n}$ is a weight vector at discrete time, $n$, 
$E_{n}$ is a generalized error, which depends on the learning procedure, and
$\epsilon$ is a learning parameter.

 Among several learning rules,
 'backpropagation' algorithm\cite{PDP},
 which is a natural extension of steepest descent method to neural networks,
 is often used for its ability 
to realize the desired mapping relation in a network.
 The algorithm is originally formulated as an on-line learning procedure. 
The on-line procedure of the backpropagation can be divided into two kinds.
The first one is a 'probabilistic on-line learning' (POL), which uses 
"local error", $E_{p_{n}}$, in Eq.(1):
$ E_{p_{n}}(\vec{X}_{p_{n}},\vec{\omega})=(\vec{\sigma}(\vec{\omega})
-\vec{Y}_{p_{n}})^{2}/2$,
where a pattern index, 
$p_{n}$, at discrete time, $n$, is drawn with pattern probability
 $P_{p}$ satisfying $ \sum_{
p=1}^L P_{p} =1$, 
 and $\vec{\sigma}$ is an output of the network.

 On the other hand, time-correlated input patterns into the network are 
often  used,
as in the case of the time-series 
on-line learning.
In such cases, the patterns may be presented in the 
deterministic order of appearance: $p_{n+1} = f (p_{n})$,
where $f$ is a map which 
produces the time series of pattern indices.
 We call this second on-line learning procedure 
as 'deterministic on-line learning'(DOL).
Although we will mainly analyze, in DOL, the case that the target function and 
the map which makes the sequence of pattern index
coincide, 
more generally one can use dynamics that is making sequences of patterns,
different from the target function.

In contrast to the on-line learning, we also discuss the 'global learning' (GL)
which is a
modified algorithm of POL.
The algorithm uses "global error", $E_{gl}(\vec{\omega})$, in Eq.(1):
 $E_{gl}(\vec{\omega})  =  \int E_{p}(\vec{X}_{p},\vec{\omega})
 \rho (p)
dp \, \, \, (p \in {\bf R})$,
 which is an averaged
error over patterns, where
$ \rho (p)$ is a probability density of the pattern with index, $p$.
The algorithm often gets easier for analysis,
 because the error does not depend on the special pattern.

 Although on-line learning does not obey exact gradient descent
process of global error as in global learning (GL),
 complete randomness of subsequent
patterns in case of POL makes 
 analytical approach possible in the context of master equations, which is 
approximated by Fokker-Planck equation in the limit of small learning
parameters\cite{Radons,Heskes,Hansen,Radons93}.
 Exactly solvable models are also discussed in the literatures\cite{Biehl,Saad}.

Recently Wiegerinck and Heskes\cite{Wiegerinck} showed theoretically that 
time-correlation between
subsequent patterns of on-line learning contributes to the diffusion 
term of a weight vector
in the Fokker-Planck 
equation approximated from the equivalent equation as Eq.(1),
and suggested that the result may help to understand the accelerated 
on-line learning with time-correlated patterns found in \cite{Mpitosos,Prog}.

  In this paper, we study how time-correlation of subsequent
patterns effects on the convergence
of learning,
by comparative studies of the two on-line learning
 procedures; a) probabilistic on-line learning  (POL) and b) deterministic
 on-line learning (DOL).
 We use the tent map in most cases as a target mapping relation, because
 the map makes this comparative study easy.
 But the result is found to be similar for other maps.
 The tent map\cite{Schuster} is written as: 
\begin{equation}
x_{n+1} = f(x_{n}) = r( 1 - 2 | x_{n} - 1/2 |) .
\end{equation}
We use sequence of patterns which is produced by the tent map itself
in DOL.
When $r=1$, the time series produced by the map has
  white Fourier spectrum and constant invariant density
between [0, 1] as same as the uniformly random number, [0, 1]; where
the deterministic nature of chaotic correlation is expected to appear
clearly in comparison with probabilistic randomness.

  Let us now consider a conventional feedforward neural network with an input, 
and output terminals and $N-2$ hidden layers with $M$ neurons.
The output of $i$-th neuron of the
$m$-th layer of the
  network is as follows:
\begin{equation}
\left\{
\begin{array}{rcl}
 y_{i}^{2} &=& \tanh(\omega_{i}^{1}x-\omega_{i0}^{1}), \\
 y_{i}^{3} &=& \tanh( \sum_{j=1}^M \omega_{ij}^{2} y_{j}^{2}  
-\omega_{i0}^{2}), \\
 \vdots & & \vdots \\
y_{i}^{N-1} &=& \tanh( \sum_{j=1}^M \omega_{ij}^{N-2} y_{j}^{3}  
-\omega_{i0}^{N-2}), \\
\sigma &=&  \sum_{i=1}^M \omega_{i}^{N-1} y_{i}^{N-1} ,  
\end{array}
\right.
\end{equation}
where $\omega^{1}_{i}$, $\omega^{2}_{ij}$, $\cdots$, $\omega^{N-1}_{i}$ 
are the synaptic weights connecting the input terminal to the second layer 
neurons, second to third layer, $\cdots$, and ($N$-1)th to output, $\sigma$, 
respectively, $\omega_{i0}^{m-1}$ is a bias term to $i$-th
 neuron of $m$-th layer.
In this paper, we restrict ourselves for simplicity
 in the case that $N=4$ and $M=3$.
 The hidden layers ($y^{2}$, $y^{3}$, $\cdots$, $y^{N-1}$) have full 
inter-layer connections.
 The local error, $E$, is written:
\begin{equation}
 E(x_{n},\vec{\omega})=(\sigma(\vec{\omega})-f(x_{n}))^{2}/2 ,
\label{eq:SQE}
\end{equation}
where $f(x)$ is the functional relationship of the tent map. 
 Global error  
is also used in on-line learning to evaluate 
how learning progresses, because global error 
does not depend on 
the special input pattern, $x_{n}$.

 It is known that learning curves 
decrease suddenly between plateaus for many target functions and models.
 In case of this tent map function learning, there usually exists 
a critical time when
 the global error, $E_{gl}$, decreases sharply, and the map learned by 
the network shifts abruptly from a constant to a tent~\cite{Prog}.
Thus, one can easily define the converging
time, $t_{cr}$, when the global error crosses the geometrical mean between 
$E_{gl}$ on the first plateau and that on the second plateau 
(see Fig.1).
 The typical learning curves of the tent map function are shown in Fig.1.
Generically, the three converging times of the tent map learning are found
 to satisfy the following 
inequality: 
$t_{cr}^{c} \le t_{cr}^{r} \le t_{cr}^{g}$,
where $t_{cr}^{c}$,  $t_{cr}^{r}$ and $t_{cr}^{g}$ are the converging times 
of DOL, POL and GL respectively.
 Notice that the invariant density, $\rho (x)$, of GL 
and that of POL are always made same as 
that of chaotic input (DOL) for comparative purpose.
 The order of three converging times is consistent with previous 
reports\cite{Mpitosos,Prog,Nakaj}.
 As one expects from the dynamical equations for weight vectors,
 the three converging
times coincide for $\epsilon \rightarrow 0$.

 How is the effect of deterministic randomness of subsequent patterns,
 which follows the  
 chaotic time series, related to that of probabilistic ones?
First we concentrate ourselves on this problem 
 to discuss the difference of converging time between DOL ((b) in Fig.1) 
and POL ((c) in Fig.1). 
Recent studies show that chaotic perturbation
has anomalous effects on complex 
systems such as Hopfield model\cite{Hayakawa} and general multi-stable
 systems\cite{Journal}, even if 
the simple statistical quantities (mean, variance, probability 
density and
Fourier spectrum) of chaos coincide with that of random noise. 
The effects are known to be related to the
unstable fixed points of chaos.
Chaotic force has transiently 
strong time-correlation when input pattern, $x$, is in the 
neighborhood of unstable fixed points; 
 these are
$x^{*}=0$ and $x^{*}=2/3$ in the tent map with $r=1$.
The nearer is the input, $x$, injected to one of the unstable fixed points, 
the longer $x$ stays in the neighborhood.
Therefore 
the network of DOL {\em sees} biased (or, special) patterns for a while 
during which the input, 
$x$, stays several times in the vicinity of the unstable 
fixed point. 
In this period, the system moves to the direction 
continuously to reduce the special 
local error,
 $E(x^{*})$, for a while, i.e. the system
is largely moved without constraint of global error due
to the unstable fixed points of the chaotic map.
This phenomenon is easily verified
by numerical simulation as in Fig.2.
 It should be noticed that
 the direction of the motion of the weight vector in this period
is not necessarily the one which reduces 
the global error, $E_{gl}$. 
On the other hand,
when the input, $x$, stays apart from an unstable fixed point,  
the sequence of input is almost as random as probabilistic; 
therefore the large change of 
weight vector in finite time steps is unlikely to occur, 
and the system is expected to move mostly along a gradient descent path 
of global error.

 The difference of time-correlation of input 
patterns  affects the convergence
of learning largely 
even when all the simple statistical quantities 
coincide between the tent map chaos and the uniform random
as mentioned before.
 Therefore, it is required to clarify the effect of this chaotic 
 time-correlation 
on the convergence of learning.
 In DOL, correlation range of 
input can be varied by the change of iteration number, $N$,
 as the selection rule of the sequence of patterns,
$x$, as $x_{n+1} = f^{N}(x_{n})$ by fixing the target function, $f$,
as $x_{n+1} = f(x_{n})$.
In the strong chaos limit, 
$N \rightarrow + \infty $, the time-correlation of the subsequent
 input, $x$, 
dissappers: the sequence of the input pattern is expected to be 
 as random as probabilistic. 
Fig.\ref{fig:b} shows that the time-correlation of weak chaotic input 
($N=1$) certainly works
 to accelerate 
time series learning. 
 Fast decay of the effect of time-correlation of 
subsequent patterns on the acceleration is observed:
 $t^{c}_{cr}$ for $N=2$ is nearly equal to that for $N=100$. This is 
 found to be consistent with the 
 exponential decay of deterministic correlation with increasing
 $N$\cite{Journal}.
 Saturated value of $t^{c}_{cr}$ is equivalent to the one given by the 
learning time for random input (POL).

The effect of the time-correlation of the input on the 
function learning decreases 
 with decreasing $\epsilon$, and it is completely 
annihilated in the adiabatic limit, $\epsilon \rightarrow 0$; 
where the change of weight vector per unit time is so small that the evolution 
of the system is shortly averaged over pattern indices\cite{Amari}. 
Therefore the dynamics of POL and that of DOL (and also GL) 
should coincide with each 
other in this limit.
Equation.(1) indicates that the continuous time, $T$, as used
in Fokker-Planck description\cite{Wiegerinck}, should be proportional
to $\epsilon$. Therefore the converging time, $t_{cr}$, in the discrete model
 both for POL and for DOL
should scale with $\epsilon^{-1}$, and $\epsilon t_{cr}$
 should be independent of
$\epsilon$ in the $\epsilon \rightarrow 0$ limit.
However, it has not been understood how the finite learning parameter, 
$\epsilon$,
affects
the accelerated learning, that is how  $\epsilon t_{cr}$
 should behave with $\epsilon$.

 One finds from the result of simulation (Fig.4) that 
the two normalized converging times, $\epsilon t_{cr}$, approach
to the same value
in the small learning parameter limit ($\epsilon \rightarrow 0$).  
 Approach of the normalized converging time to finite value in the
 limit
shows that there is no local minimum in the
learning process.
 If there are any local minima in the learning process, the
normalized converging time must diverge to infinity as $\epsilon
\rightarrow 0$\cite{Kushner}.
In POL, the normalized converging time 
increases monotonically with increase of the learning parameter, $\epsilon$.
However, in DOL, the normalized converging time, $\epsilon t_{cr}^{c}$,
decreases first with increase of $\epsilon$, and after some learning parameter,
$\epsilon_{opt}$, it increases monotonically.

As known in general relaxation methods, finite stepping parameter, $\epsilon$,
is harmful, because
the possibility of overshooting in phase space increases as $\epsilon$ 
increases. Therefore, the normalized converging time is expected to 
increase monotonically with increasing learning parameter\cite{Saad} as 
the result
of overshooting in a 
learning process without local minima.
The simulation showed that this is the case for POL but not necessarily 
for DOL (Fig.4). 
The decrease of $\epsilon t_{cr}^{r}$ with increase of $\epsilon$ was not
observed in the simulations (Fig.4).
On the other hand, decrease of $\epsilon t_{cr}^{c}$ was often found
 in chaotic patterns, not only
in the learning of the tent map but also in that of
logistic map with several parameters.

As one notices, the reduction of converging time with increase of the 
learning parameter
 is possible when the system has to escape from local
minima to reach the solution of learning\cite{Kushner}. But the present 
system  has no local minima.
One might think it strange that the normalized converging time decreases
as the learning parameters increases in a process without local
 minima.
There should be an alternative which overcomes the harm
of overshooting in the region $ 0 < \epsilon < \epsilon_{opt}$ in DOL.

We found that  the puzzle may be solved by 
noticing the fact that there are generically in learning process
plural gradient descent
paths to the solution. 
If chaotic correlation of subsequent patterns works effectively 
to find a shorter path to the solution by its diffusive motion of weight space,
 the observed phenomena are understandable.
The possibility is strengthened by the
fact that the system under chaotic patterns (DOL)
should be largely moved away from the exact gradient
descent direction of global error due to the unstable fixed points,
 which would facilitate the system to cross
 over the potential barrier
between the gradient descent paths.

The same order of diffusive motion against gradient descent
direction of global error, as found in DOL (see, Fig.1), 
would be possible, in principle, even in POL, with
 larger learning parameter, $\epsilon$.
However, increase of $\epsilon$ strengthes the harm of overshooting 
simultaneously: the harm of overshooting may cancel the merit 
the diffusive motion in POL. 
In DOL, the harm of overshooting overcomes the
merit of the diffusive motion when $\epsilon$ go over
$\epsilon_{opt}$ where the normalized converging time begins to increase.

Finally, we mention an automatic reduction mechanism of the fluctuation
of the system, which is
characteristic of on-line learning and may weaken the harm of overshooting
with a finite learning parameter. 
As discussed in some literatures\cite{Radons,Heskes},
in "perfectly trainable
networks"\cite{Funahashi}, in which $E_{gl} (\vec{\omega})=0$ is available, 
the fluctuation in weight vector space (equivalently,  the diffusion
rate in Fokker-Planck representation \cite{Wiegerinck}) becomes zero when 
the system reaches error-free ($E_{gl} = 0$) state:
 the error-free state behaves as a "sink" of probability
flow\cite{Heskes}.
The reduction of the fluctuation can also occur even if the network is not
"perfectly trainable": the system should be stabilized when the residual
error is small enough\cite{Last}.

We showed in this paper that
the accelerated on-line learning with chaotic patterns is 
attributed to the unexpected scaling of the converging time with
learning parameter, $\epsilon$:  
 the converging time, $t_{cr}$, decreases much faster than $t_{cr} \approx
\epsilon^{-1}$ with increasing $\epsilon$ even without local minima. 
The results may indicate the beneficial aspects of finite learning parameters
of on-line learning with time-correlated patterns, because in any case
one is forced to use finite learning parameters
in realistic learning processes.
The studies of the optimal time-correlation of general patterns and/or the 
optimal
learning parameter for the network,
 together with the proof of acceleration mechanism,
are under way.

\begin{figure}[b]
\caption{Typical learning curves of the tent map function by three learning
methods:
a) global learning, b) deterministic on-line learning,
c) probabilistic on-line learning.
Invariant density of a) and c) are made the same as b).
The initial conditions of the weight vectors are the same,
 $\epsilon = 0.05$  and $r=0.95$.
}
\label{fig:lc}
\end{figure}

\begin{figure}[b]
\caption{(a) Typical temporal evolution of input, $x$, found in DOL,
where $ t < t^{c}_{cr}$.
(b) Corresponding time evolution of averaged velocities of a weight vector,
$\vec{\omega}$, in a finite time interval, T,
 where $\delta \omega \equiv |\delta \vec{\omega}|
 = | \vec{\omega}_{n+T/2} - \vec{\omega}_{n-T/2} | $.
Initial value of $\vec{\omega}$ is drawn from uniformly random number
 [-0.05, 0.05],  $\epsilon = 0.05$ and  $r=0.9995$.}
\label{fig:a}
\end{figure}

\begin{figure}[b]
\caption{Converging time, $t^{c}_{cr}$, versus several deterministic
 time-correlation of input patterns,
 $x$. Lyapunov exponent, $\lambda$, of the sequence of input is:
$\lambda = N \log 2$.
 In the (strong chaos) limit, $N \rightarrow \infty$,
the system is almost equivalent to POL with
uniformly random input $[0, 1]$.
Ensemble averages over 100 initials are shown. Initial
value of $\vec{\omega}$ is drawn from  uniformly random number $[-0.05, 0.05]$,
$\epsilon = 0.05$ and $r=0.9995$.
It is found that $t^{c}_{cr}(N \gg 1)$  $\approx t^{r}_{cr}$ (POL).}
\label{fig:b}
\label{fi:sc1}
\end{figure}

\begin{figure}[b]
\caption{Dependence of normalized converging time, $\epsilon \, t_{cr}$,
on learning parameter, $\epsilon$,
 for the tent map learning (solid line for DOL, dotted line for POL).
Ensemble averages over 100 initials are shown.
Initial value of $\vec{\omega}$ is drawn from uniformly random number
[-0.1, 0.1] and $r=0.9995$.
}
\label{fig:c}
\label{fi:sc2}
\end{figure}

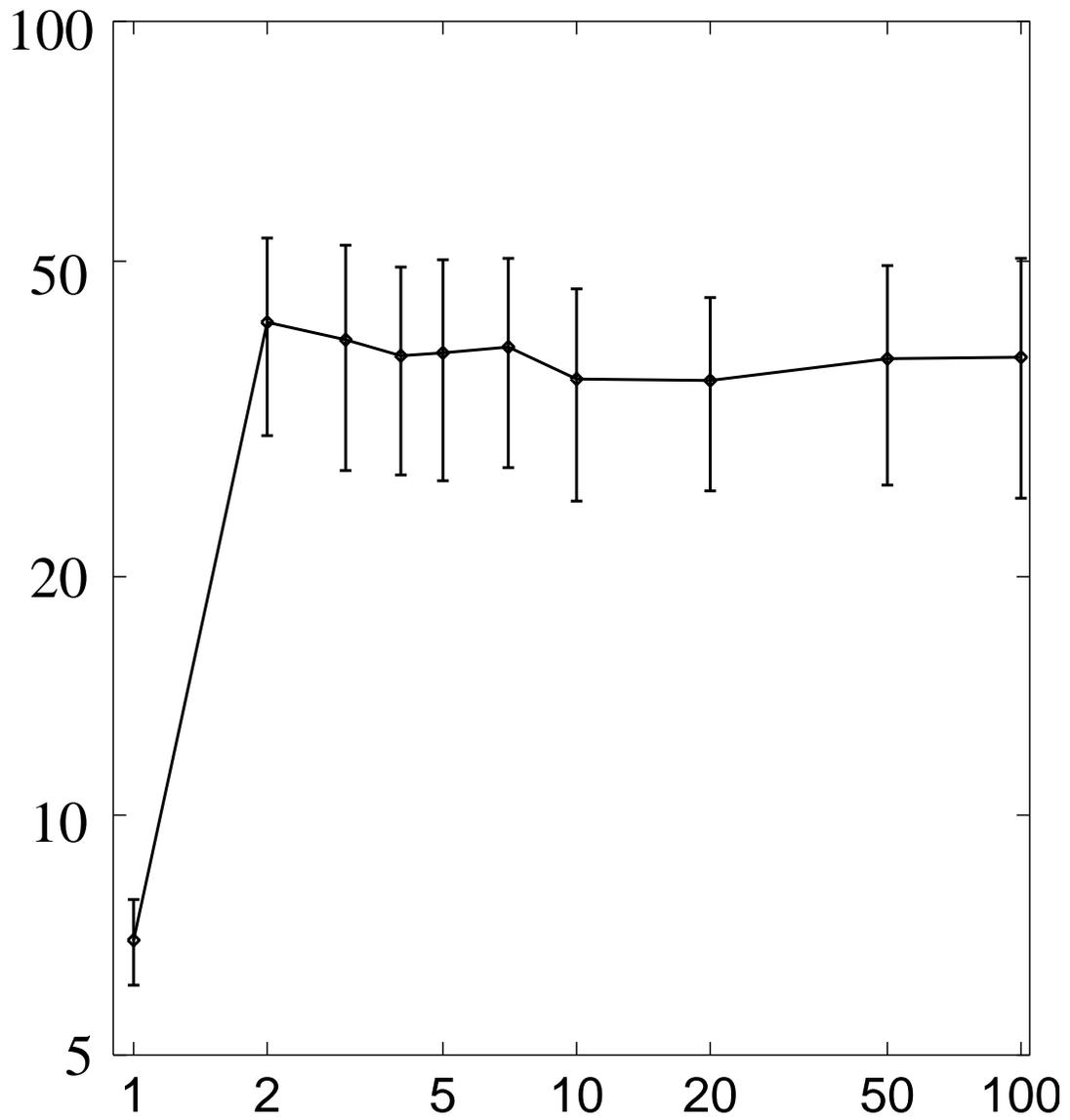
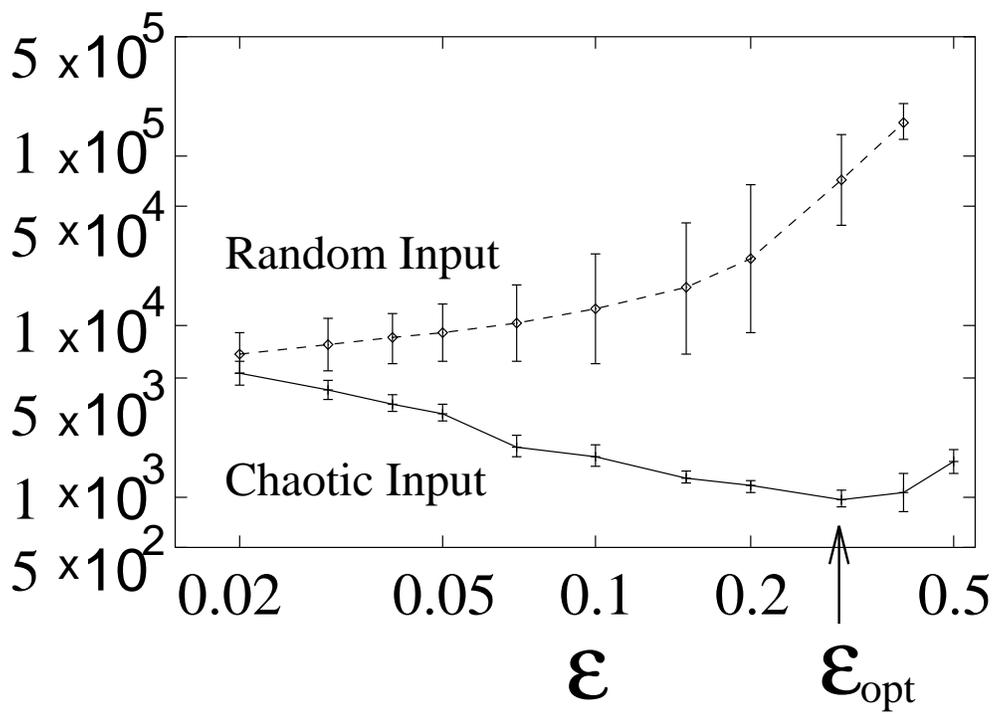
\end{document}